\documentclass[10pt,twocolumn]{article} 

\usepackage{simpleConference}
\usepackage{times}
\usepackage{graphicx}
\usepackage{amssymb}
\usepackage{url,hyperref}

\usepackage{graphicx}%
\usepackage{multirow}%
\usepackage{amsmath,amssymb,amsfonts}%
\usepackage{amsthm}%
\usepackage{mathrsfs}%
\usepackage[title]{appendix}%
\usepackage{xcolor}%
\usepackage{textcomp}%
\usepackage{manyfoot}%
\usepackage{booktabs}%
\usepackage{algorithm}%
\usepackage{algorithmicx}%
\usepackage{algpseudocode}%
\usepackage{listings}%
\usepackage{caption}
\usepackage{bbding}
\usepackage{subcaption}
\usepackage[bottom]{footmisc}
\usepackage{xcolor}
\definecolor{RossoRubino}{RGB}{193,71,50}

\definecolor{DragonGreen}{RGB}{0,126,48}

\definecolor{CrystalBlue}{RGB}{12,43,193}

\begin{document}

\title{HydraScreen: A Generalizable Structure-Based Deep Learning Approach to Drug Discovery}

\author{Alvaro Prat*, Hisham Abdel Aty, Gintautas Kamuntavičius, \\
Tanya Paquet, Povilas Norvaišas, Piero Gasparotto, Roy Tal \\
\\
AI Chemistry, Ro5 \\ 2801 Gateway Drive, Irving, 75063, TX, USA \\
\today
\\
\\
*Corresponding author. E-mail: \url{aprat@ro5.ai}  \\
}

\maketitle
\thispagestyle{empty}

\abstract{
    We propose HydraScreen, a deep-learning approach that aims to provide a framework for more robust machine-learning-accelerated drug discovery.
    HydraScreen utilizes a state-of-the-art 3$D$ convolutional neural network, designed for the effective representation of molecular structures and interactions in protein-ligand binding.
    We design an end-to-end pipeline for high-throughput screening and lead optimization, targeting applications in structure-based drug design.
    We assess our approach using established public benchmarks based on the CASF 2016 core set, achieving top-tier results in affinity and pose prediction (Pearson's $r = 0.86$ \& RMSE $= 1.15$, Top-1 $= 0.95$).
    Furthermore, we utilize a novel interaction profiling approach to identify potential biases in the model and dataset to boost interpretability and support the unbiased nature of our method.
    Finally, we showcase HydraScreen's capacity to generalize across unseen proteins and ligands, offering directions for future development of robust machine learning scoring functions.
    HydraScreen\footnote{Accessible at \href{https://hydrascreen.ro5.ai}{https://hydrascreen.ro5.ai}} provides a user-friendly GUI and a public API, facilitating easy assessment of individual protein–ligand complexes.
}

\section{Introduction}

The drug discovery process is notoriously laborious, time-consuming, and costly \cite{hughesPrinciplesEarlyDrug2011, paulHowImproveProductivity2010}. The average cost of developing a new drug is estimated to be around \$2.6 billion, with a limited success rate below 10\% \cite{paulHowImproveProductivity2010, sunWhy90Clinical2022}. In light of this, significant efforts in the field of computer-aided drug discovery (CADD) have been pursued, with the aim of developing faster and more robust computational methods to accelerate and reduce risk behind drug discovery processes ~\cite{liStructureawareInteractiveGraph2021, musil2021physics, salentin2015plip, trottAutoDockVinaImproving2010b, wangPDBbindDatabaseCollection2004a, wangPDBbindDatabaseMethodologies2005a, wojcikowskiDevelopmentProteinLigand2019, zhengOnionNetMultipleLayerIntermolecularContactBased2019}. In particular, structure-based drug discovery (SBDD) methods have been widely used to identify and optimize drug candidates~\cite{batoolStructureBasedDrugDiscovery2019}. Such methods leverage the 3$D$ structure of both the target protein and the ligand to assess the likelihood of binding - an early indicator in the drug discovery process to identify promising drug candidates~\cite{meliScoringFunctionsProteinLigand2022b}.

On the other hand, Quantitative Structure-Activity Relationship (QSAR) models predict biological activity from molecular features and/or structural properties ~\cite{muratovQSARBorders2020, banBestPracticesComputerAided2017}. These models range from so-called traditional methods employing linear regression to non-linear counterparts using polynomial regression and artificial neural networks (ANNs)~\cite{soaresReEvolutionQuantitative2022}. In contrast, most SBDD models are able to rank and prioritize vast compound libraries using energy calculations determined by specific force fields (FFs). While QSAR models correlate molecular structures with biological activities, structure-based methods like docking are therefore able to simulate the actual interactions between proteins and ligands~\cite{trottAutoDockVinaImproving2010b}, often providing further explainability and accuracy.

Despite the availability of more precise approaches like \emph{ab initio}~\cite{antony2011protein,antony2012fully,ryde2016ligand} and free energy methods~\cite{bussi2015free,fidelak2010free,deflorian2020accurate}, docking is often used favourably at the early stages due to its computational efficiency~\cite{ferreiraMolecularDockingStructureBased2015, pinziMolecularDockingShifting2019}. This method simplifies a dynamic interaction using an empirical scoring function, which serves dual purposes: querying potential binding conformations (docking power) and predicting protein-ligand binding affinity (scoring power). The accurate prediction of binding affinity is crucial, influencing the selection of promising drug candidates~\cite{kairysBindingAffinityDrug2019}. Simultaneously, predicting the correct binding mode also affects the robustness of these models, by filtering out predictions based on implausible and unstable conformations. Besides these core applications, physics-based scoring functions are used in docking to screen compounds from high-throughput screening (HTS) databases, further cementing their role in drug discovery pipelines~\cite{seifertVirtualHighthroughputSilico2003}.

Traditionally, scoring functions are either physics-based, empirically regressed, or a combination of both. More recently, with the greater availability of assay and protein-ligand interaction data (growing at a rate of 15\% per annum \cite{btu626}), other paradigms of scoring functions have been developed, such as knowledge-based and machine-learning (ML) based scoring functions (MLSFs)~\cite{halgrenGlideNewApproach2004a, jonesDevelopmentValidationGenetic1997a, liuClassificationCurrentScoring2015, morrisAutoDock4AutoDockTools4Automated2009b, ravindranathAutoDockFRAdvancesProteinLigand2015a}.

Classical computational methods such as docking and Molecular Dynamics (MD) simulations face numerous challenges, including limitations in pose sampling and uncertainty in generating and predicting adequate binding conformations, as well as computationally scaling limitations. ~\cite{rossi2016anharmonic,durrantMolecularDynamicsSimulations2011,eberhardtAutoDockVinaNew2021, morrisAutoDock4AutoDockTools4Automated2009b, trottAutoDockVinaImproving2010b}.
Moreover, existing benchmarks and evaluation protocols often inadequately represent real-case scenarios, leading to biased results and limiting the applicability of these methods~\cite{chenHiddenBiasDUDE2019a, francoeurThreeDimensionalConvolutionalNeural2020a, siegNeedBiasControl2019a, wallachMostLigandBasedClassification2018b}.

Structure-based binding affinity methods hold promise for addressing these issues by learning unbiased scoring functions directly from the ever-increasing pool of 3$D$ protein-ligand complexes annotated with experimentally obtained affinity labels.
This is now possible due to the large influx of new high-quality structural data available as a result of advancements in experimental techniques such as X-ray crystallography~\cite{leonarski2018fast,weinert2017serial,nogly2018retinal,skopintsev2020femtosecond}, nuclear magnetic resonance and cryo-electron microscopy~\cite{congreveStructuralBiologyDrug2005}. However, the performance of these methods is often limited by the quality of the training data, which is frequently biased and imbalanced. Moreover, the lack of robust benchmarks and evaluation metrics further hinders the development of accurate prediction models~ \cite{chenHiddenBiasDUDE2019a, francoeurThreeDimensionalConvolutionalNeural2020a, siegNeedBiasControl2019a, wallachMostLigandBasedClassification2018b}. Through learning these interactions, however, models can internalise an unbiased scoring function, enhancing the novelty and diversity of screened compounds (in contrast to filtering out compounds selected with a biased approach). Additionally, rapid identification and removal of suboptimal compounds encourage faster drug discovery cycles, paving the way for efficient large-scale HTS.

Classical MLSFs have been used widely over the last decade, with MLSFs consisting of standard ML foundation models including Support Vector Machines (SVM), Random Forest (RF), and Na\"ive Bayes classifiers, to predict protein-ligand binding sites or affinities~\cite{meliScoringFunctionsProteinLigand2022b}. More recently, with increase data availability, deep learning (DL) methodologies have taken preference. For instance, AtomNet~\cite{wallach2015atomnet} and KDEEP~\cite{jimenezKDEEPProteinLigand2018a} are pioneering attempts at using Convolutional Neural Networks (CNNs) to predict protein-ligand affinity. Similarly, OnionNet~\cite{wangOnionNet2ConvolutionalNeural2021, zhengOnionNetMultipleLayerIntermolecularContactBased2019}, presented an improved CNN-based method by considering multi-layered intermolecular contacts. Additionally, ensemble-based and Graph Neural Network (GNN) strategies have been employed by methods like AK-score~\cite{kwonAKScoreAccurateProteinLigand2020a} and graphDelta~\cite{karlovGraphDeltaMPNNScoring2020}, respectively.

Whilst these upcoming solutions have demonstrated promising results, they are often solely trained as scoring functions over co-crystal complexes \cite{wangOnionNet2ConvolutionalNeural2021, zhengOnionNetMultipleLayerIntermolecularContactBased2019, jimenezKDEEPProteinLigand2018a, karlovGraphDeltaMPNNScoring2020, kwonAKScoreAccurateProteinLigand2020a}. Not only is crystallizing protein-ligand complexes prohibitively expensive for early-stage drug discovery, making models trained on crystal data unrealistic for practical scenarios, but it also strips any notion behind the entanglement between successful protein-ligand binding (pose estimation) and affinity prediction. 

When designing robust MLSF solutions, it is important to consider some of the new challenges that arise with the introduction of ML methodologies. Particularly, it has been well established in the literature that a significant portion of structure-based MLSFs are heavily biased towards ligand properties and features, rather than the corresponding protein-ligand interactions \cite{chenHiddenBiasDUDE2019a, francoeurThreeDimensionalConvolutionalNeural2020a, siegNeedBiasControl2019a, wallachMostLigandBasedClassification2018b}. This bias often results in the frequent selection of promiscuous ligands for unseen targets, therefore selecting unfeasible drug candidates. Moreover, the performance of deep learning models often reflects the distribution of the training dataset, therefore evaluation datasets lacking representativeness or diversity can confine the models' generalizability, yielding poor performance when applied to out-of-domain data. 

HydraScreen is a DL model devised to alleviate the drawbacks of prevailing computational approaches by accurately predicting protein-ligand binding affinity and pose estimation with top-tier precision, matching extant state-of-the-art methods. 

In this study we detail our proposed methodology to obtain robust structure-based MLSFs, delving into the architecture, data treatment, and datasets used to train and evaluate HydraScreen. We introduce a curated open-source dataset, RD-2020, and show robust performance across a series of dataset partitionings based on similarity~\cite{suTappingBlackBox2020a} and temporal splits, addressing the limitations of existing overly optimistic benchmarks for protein-ligand modelling. We evaluate the limitations of current datasets and outline a novel approach for visualizing protein-ligand landscapes correlated with various types of annotations, demonstrating HydraScreen's unbiased performance. We also introduce the Protein-Ligand Interaction Ensemble (PLIE) score as a proxy for quantifying the quality of docked pose ensembles and validate its effectiveness across a series of metrics of interest. Last, we summarise our key considerations for future works and provide an open-source gateway to use HydraScreen across different stages of drug discovery.

\section{Methods}
In this section, we present the model architecture and framework proposed for building and deploying robust MLSFs, followed by an introduction to the datasets used to train and evaluate Hydrascreen. We introduce our open-source dataset, RD-2020, and present our preferred approach to evaluate our model's generalisation ability, reliability and explainability.

\subsection{HydraScreen}
\label{methods-hydrascreen}
\begin{figure*}[ht!]
    \centering
    \includegraphics[width=0.75\textwidth]{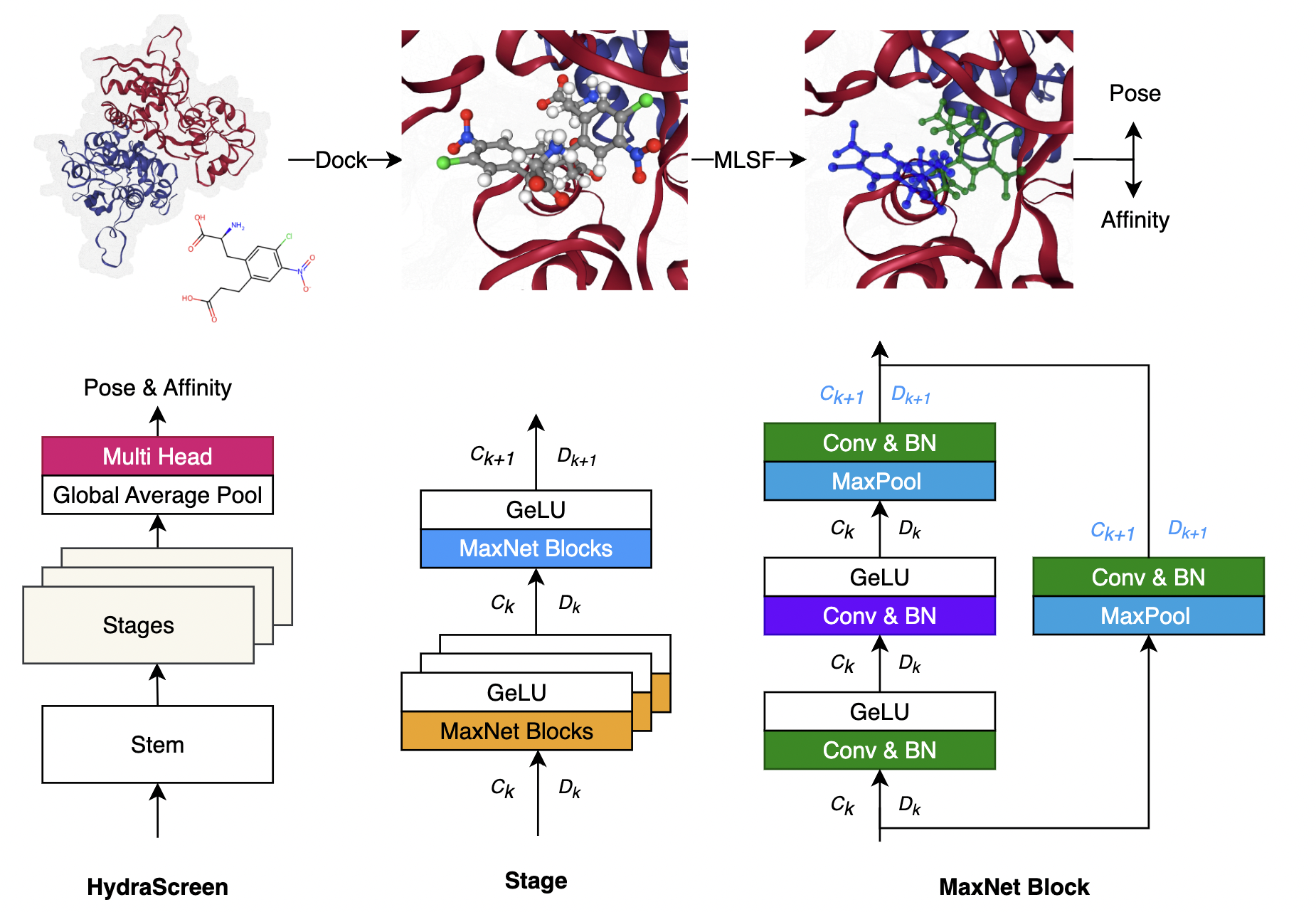}
    \caption{Structure-based screening framework (top): A query molecule is docked against the target protein of interest, creating an ensemble of poses in the pocket, used as inputs to an MLSF. On the right, a pose with high RMSD (blue) and the native crystal pose of that same ligand (green) are illustrated; HydraScreen architecture (bottom): 3$D$ CNN model composed of 3 main sections: (i) A stem which performs an initial grid reduction convolution (larger strided kernel) from a featurized protein-ligand complex; (ii) A series of stages composed of an array of MaxNet blocks. Each MaxNet block is a 3$D$ residual network optimized for increased receptive field and reduced parameter count: shortcut and tail cells are built with kernel size 1 (green) and core cells with kernel size 3 and grouped convolutions (purple). Note that only the last block in each stage performs a spatial reducing MaxPool operation as well as channel expansion (blue). $C_k$ and $D_k$ respectively represent the number of channels and spatial dimensions of the feature maps in the $k$'th layer; (iii) A flattening 3$D$ spatial average pooling followed by multiple multi-layer perceptron networks (for each output).}
    \label{fig:hydrascreene2e}
\end{figure*}

We create an end-to-end framework for SBDD in a twofold manner. First, we leverage docking to generate poses in the pocket of the target protein. Subsequently, sampled protein-ligand pose ensembles are featurized and scored according to our MLSF (Figure \ref{fig:hydrascreene2e}). 

\textbf{Generating Docked Ensembles.} Docked ensembles are generated by docking the query ligand into the target's pocket with Smina~\cite{koesLessonsLearnedEmpirical2013a}. Before docking, water molecules, metal ions, and redundant co-crysallized ligands are removed from the protein PDB file using the ADFR Suite~\cite{ravindranathAutoDockFRAdvancesProteinLigand2015a}. Subsequently, sample ligand conformations are generated as seeds via RDKit~\cite{landrum2013rdkit} and processed with Meeko \cite{MeekoPreparationSmall2023}. In order to generate meaningful poses we constrain the search space within a grid defining the target's pocket using the autobox option. We extract at most 20 poses for each docking ensemble with at a minimum distance of 0.5~\AA{} Root Mean Square Displacement (RMSD) from each other. Note that in this study, we use a symmetry-corrected RMSD \cite{Meli2020} calculation. When present, we define the pocket by keeping faces 4~\AA{} away from the furthest atom in the $\mathrm{E}(3)$ aligned reference pose (crystal or energy minimized native pose), as well as including it in the docked ensemble. 

\textbf{Featurization.} The inputs to HydraScreen are prepared by meshing the protein-ligand complex into a 3$D$ cubic grid using libmolgrid~\cite{sunseri2020libmolgrid}, a CUDA accelerated library for generating molecular grids. We use a 24~\AA{} grid centered at the center of mass of the ligand, with a 0.5~\AA{} resolution ($D_0 = 49$). Ligand and protein atoms are pre-processed in a two-step process.
First, atoms are mapped into distinct features using a reduced set from the Openbabel Atom Typer reference set~\cite{oboyleOpenBabelOpen2011}, producing $f_p$ and $f_l$ distinct protein and ligand features respectively. A more detailed description of the atom feature types used is provided in the SI.

Second, continuous densities are sampled at each grid-point using an isotropic Gaussian kernel. A separate channel is made for each atomic feature $f_i \in \{f_p, f_l\}$, resulting in a cubic grid of the protein-ligand complex $X_{pl} \in \mathbb{R}^{D_0^3(f_p + f_l)}$. Densities $\delta_i(\cdot)$ are sampled at each grid point $\textbf{x} \in \mathbb{R}^3$ as the sum of the contributions of identical atomic species with atomic centers $\textbf{c} \in \mathbb{R}^3$ (Eq.~\ref{eq:rbf}) with standard deviation $\sigma = 1/2$.

\begin{equation}
    \delta_i(\textbf{x}) \propto \sum_{\textbf{c} \in f_i} \exp\left(-\frac{{\|\textbf{x} - \textbf{c}\|^2}}{{2\sigma^2}}\right)
    \label{eq:rbf}
    \vspace{8pt}
\end{equation}

\textbf{Architecture.}
In contrast to recent efforts to create SO(3) equivariant graphs or point transformers 
\cite{thomasTensorFieldNetworks2018, weiler3DSteerableCNNs2018, kondorClebschGordanNetsFully2018, fuchsSETransformers3D2020, geigerE3nnEuclideanNeural2022,pozdnyakov2023smooth,bigi2023wigner}, we leverage a carefully designed invariant 3$D$ CNN-based architecture. Despite the clear benefits of designing 
an architecture equivariant under continuous 3$D$ roto-translations,
we believe that exploiting modern CNNs layers can still offer significant advantages. Particularly, locality and parameter sharing in convolutional filters can help capture the continuous environments present in neighbouring atomic interactions. Moreover, the hierarchical nature of CNN layers aligns well with the hierarchical nature of molecular structures. Last, translational and rotational invariance can be approximated via stacked spatial MaxPooling operations. We build our architecture and training/testing framework with Pytorch Lightning \cite{Falcon_PyTorch_Lightning_2019}, a library that provides a high-level interface for PyTorch and simplifies training and deploying DL models at scale.



\textbf{Scoring Functions.} HydraScreen returns predicted pose confidence and affinity scores for each pose within the docked ensemble. This will not only be useful to discard decoys during inference, but it will also support the model to exploit multi-modality during training, and intrinsically entangle the otherwise hidden relationship between the interactions which govern protein-ligand binding (good pose), and the interactions which discriminate between high and low ligand affinity. 

Pose score $s_p \in (0,1)$ is a proxy for how close a pose approximates to the crystallised native ligand for that target. Formally, in contrast to the common discontinuous definition commonly used in literature \cite{francoeurThreeDimensionalConvolutionalNeural2020a} we define $E_p(\cdot)$ as a smooth $\mathbb{C}^2(\cdot)$ envelope (Eq. \ref{eq:smooth_envelope}), where $r$ is the RMSD between the query and reference poses, $t$ is the desired threshold (set to 2~\AA{}), and $s$ corresponds to the squeeze factor of the envelope. A visual of the latter is shown in the SI. We use $E_p(\cdot)$ to define a cross-entropy loss (Eq. \ref{eq:ce_loss}), where $s_p^*$ is the prediction of the model wrapped by a sigmoid function.

\begin{equation}
    \label{eq:smooth_envelope}
    E_p(r) = \frac{1}{1 + e^{-s(t - r)}}
    \vspace{8pt}
\end{equation}
\begin{equation}
    \label{eq:ce_loss}
    \begin{split}
    \mathcal{L}_{p}(r, s_p^*) = -E_p(r)\log(s_p^*) \\ - \hspace{4pt} (1 - E_p(r))\log(1 - s_p^*)
    \end{split}
\end{equation}

In addition to our efforts in creating a smoother envelope, we also include the raw RMSDs as part of HydraScreen's multi-head output and regress to it during training (Eq. \ref{eq:rmsd_loss}), where $r^*$ represents the predicted RMSD between the true and the query pose. We believe this can provide additional gradient signals to help improve discrimination amongst poses with different RMSDs, instead of, for instance, labelling poses with RMSDs of 3\AA{} in the same fashion as one with 8~\AA{} (resulting in information loss). 

\begin{equation}
    \label{eq:rmsd_loss}
    \mathcal{L}_r(r, r^*) = \|r - r^*\|^2
    \vspace{8pt}
\end{equation}

 Affinity score $s_a \in \mathbb{R+}$ is directly indicative of how potent the model believes the query ligand to be, whilst assuming that particular conformation.  We set up a reinterpretation of a continuous hinged loss based on the aforementioned RMSD envelope (Eq. \ref{eq:hinge_affinity}), where $s_a^*$ is the predicted affinity of the protein-ligand complex, and $JS$ is the Jensen–Shannon divergence. The latter penalises distribution shift between the true and predicted affinity distributions and was added as a regularisation term in an attempt to alleviate the otherwise mean-seeking behaviour.

\begin{equation}
    \label{eq:hinge_affinity}
    \mathcal{L}_a(r, s_a, s_a^*) = E_p(r)\big(\|s_a - s_a^*\|^2 + JS(s_a || s_a^*)\big)
    \vspace{8pt}
\end{equation}

\textbf{Interaction Ensembles.} Within virtual screening campaigns, MLSFs and other traditional SBDD solutions typically take the best-predicted pose of a docked ensemble either by means of an independent pose scoring function or by selecting the lowest energy conformation. We offer a different solution by exploiting the predicted landscape of the pose ensemble, with the aim of extracting more information than with a single sample. Specifically, we create 
the Protein-Ligand Interaction Ensemble (PLIE) score (Eq. \ref{eq:plie}, where $\beta$ corresponds to a weighting factor) with the aim of quantifying the "quality" of the docked ensemble, an attribute that is often criticised in industry. In a similar fashion, by substituting the predicted pose score $s_p^*$ with the affinity score $s_a^*$, we compute the overall predicted affinity of a docked ensemble. 

\begin{equation}
    \label{eq:plie}
    PLIE = \sum_{ens} \frac{s_p^* e^{\beta s_{p}}}{k}; \hspace{6pt} k = \sum_{ens}\limits e^{\beta s_{p}}
    \vspace{8pt}
\end{equation}

\textbf{Training.} We train our models by aggregating the aforementioned loss functions and minimizing them via gradient descent. We use the AdamW optimizer \cite{loshchilovDecoupledWeightDecay2019a} with a batch size of 256 and a warm-up learning rate scheduler with initial, peak, and final learning rates of $10^{-6}$, $3\times10^{-4}$, and $10^{-5}$. Early stopping is also implemented to halt training by extracting an independent evaluation set from the train set. For each dataset we train an ensemble of models in parallel across two A100 GPUs, using PyTorch's native amp accelerated half-precision training. On average, HydraScreen converges in around 200,000 steps, taking around nine hours to train a single model. 

\subsection{Datasets}
We consider the current landscape of public datasets for learning MLSFs and their limitations, and introduce our curated augmented datasets, which aim to address these shortcomings. The two main dataset sources of binding affinity used in literature and in this work are the PDBbind 2020 and PDBbind 2016 sets, which compose of manually curated protein-ligand crystal structures from the PDB, with their respective experimental labelled affinity either as dissociation constants Ki, Kd, or inhibition constant IC50. \cite{liuForgingBasisDeveloping2017a, wangPDBbindDatabaseCollection2004a, wangPDBbindDatabaseMethodologies2005a}. The authors of PDBbind further subdivide these sets into refined and general sets. The refined set was developed with threshold limits to determine structure quality. For example, only ligands lighter than 1000 Da, with Ki or Kd measurements and are part of a crystal structure with a resolution greater than 2.5\AA{} are included. 

From the PDBbind 2016 refined set, comes the Comparative Assessment of Scoring Functions (CASF) benchmark, consisting of 285 protein-ligand complexes selected for protein and binding affinity diversity, and is often used as the gold-standard for benchmarking affinity prediction. The benchmark is clustered by target similarity (57 targets) and is designed to assess four major protein-ligand predictive tasks: scoring power, ranking power, docking power and screening power.

\textbf{RD-2020.} We aggregate protein-ligand complexes from both PDBbind refined and general splits to introduce RD-2020, an open-source curated dataset of over 290,000 protein-ligand poses, which aims to address some of the limitations of existing bias in datasets by: (i) releasing diverse decoy poses for each co-crystallised complex; (ii) defining a richer and more realistic benchmark for evaluating of MLSFs.

RD-2020 is filtered and augmented through extensive clean-up and docking procedures. Initially, we take all 19,443 protein-ligand complexes from PDBbind 2020 and the 65 complexes from PDBbind 2016 that are not present in PDBbind 2020. We then consider the dock-ability of the complexes based on a set of criteria including the presence of binding co-factors, resolution and nature of the crystal structure, steric clashes within the system, as well as the proposed system pH. Note that $\sim $13\% of the original ligands were not readable and therefore were not re-docked. Moreover, large polymer ligands, and structures extracted via NMR were also omitted. However, structures that were not re-docked were kept in the dataset. Finally, we use our proposed docking pipeline, at scale, to generate up to 20 poses for each selected complex by re-docking the native bound ligand into its pocket.

\label{sec:datasets}
\textbf{Generalisation I: Similarity Splits}. 
We study the effect of reducing possible existing biases in the training set by filtering complexes in the Refined 2020 set (train) by their similarity to complexes in the CASF-16 core set (test), as described in \cite{suTappingBlackBox2020a}. Similarities between the two sets can be referred to as soft overlaps, where complexes are not identical but are similar enough in terms of ligand and binding site similarity that they could artificially inflate the power of a given MLSF through overfitting. Non-redundant training sets are created based on three similarity metrics: (i) Protein sequence similarity; (ii) Ligand shape similarity; (iii) Binding site similarity. If the aggregate similarity score between a complex in the refined set and any complex in the core set is above a certain threshold, then the complex is removed from the training set. We use similarity thresholds of 0.8, 0.85, 0.9 and 0.95. Assessing performance in this setup should give better insights towards the generalizability of HydraScreen and its performance on unseen targets and ligands.


\textbf{Generalisation II: Temporal Split.} Creating a de-biased test set for MLSFs is a very difficult and imperfect task, as generalisation is rather subjective and is inherently very task-specific: users might want to exploit their MLSFs for high-throughput screening of a compound library over a target that is not present in the training set, and some users might want to perform lead optimisation within an unseen target protein. In light of this, we create a held-out set, extracting PDBs from the temporal difference between PDBbind 2020 and 2016 sets, rather than taking subsets of the same dataset, as is the case for the CASF benchmark.

\subsection{Visualising the structural landscape of protein-ligand complexes}
\label{sec:methods-soaps}

Understanding and visualizing the intricate structural landscape of protein-ligand complexes presents significant challenges. To streamline this process, we developed an agnostic structural analysis that converts the 3N all-atom coordinates of each complex into a high-dimensional representation using the Smooth Overlap of Atomic Positions (SOAP). Introduced by Bartok et al.\cite{bartok2013representing}, SOAP vectors adeptly encapsulate radial and angular structural correlations surrounding individual atoms into a high-dimensional vector that can be used to measure the similarity between atomic environments (refer to SI for expanded details)~\cite{bartok2013representing,de2016comparing,willatt2019atom,darby2022compressing}.

\begin{figure}[ht!]
\centering
\includegraphics[width=0.46\textwidth]{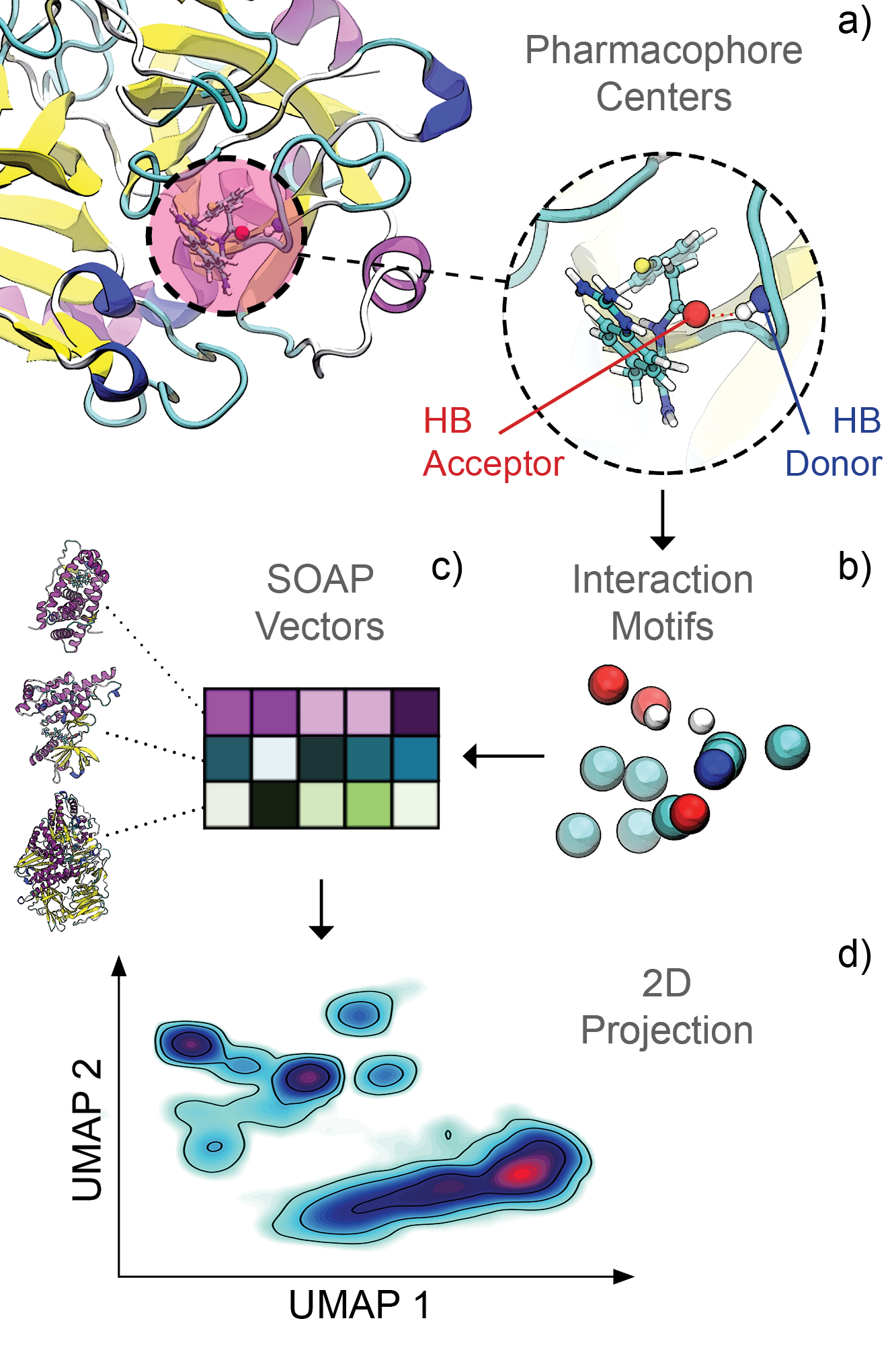}
\caption{Schematic summarizing the end-to-end process of transforming the all-atoms 3N coordinates of protein-ligand complexes into a 2$D$ SOAP projection. (a) Identification of pharmacophore centers using PLIP. (b) Approximation of each protein-ligand complex to its primary interaction motifs, focusing on the pharmacophore centers. (c) Computation of high-dimensional embeddings for each complex using SOAP vectors. (d) 2$D$ representation generation after dimensionality reduction of SOAP vectors with UMAP.}
\label{fig:plipsoapschema}
\end{figure}

Given the extensive structure of protein-ligand complexes and our primary interest in discerning specific interaction motifs between the ligand and the protein's pocket that influence activity prediction, we first simply the structure into the set of interaction patterns between the ligand and protein. This is achieved using the protein-ligand Interaction Profiler (PLIP)~\cite{salentin2015plip}, which lists for each complex the interactions between the ligand and protein and the pharmacophore centers. By selecting only the pharmacophore centers, we approximate the protein-ligand complex into a concise 3$D$ model that encapsulates only those structural motifs that are crucial for the binding. We then proceed to compute SOAPs for each pharmacophore center in each complex using \texttt{librascal} (we use the default hyperparameters and set an interaction cutoff distance of 6.0, while maximum radial and angular values are set to 5 and 6 respectively)~\cite{musil2021efficient}. Subsequently, by averaging all SOAP vectors within a complex, we obtain a unified global vector that describes the entire complex. A schematic of our approach is shown in Figure \ref{fig:plipsoapschema}.

To reduce the dimensionality of the SOAP vectors, we first employ the CUR decomposition~\cite{imbalzano2018automatic}, reducing the SOAP vectors from 9,625 to 400 dimensions -- while retaining 98\% of the explained variance. This is followed by a subsequent application of the Uniform Manifold Approximation and Projection (UMAP)~\cite{mcinnes2018umap} method (with the number of neighbours set to 10 and a minimum distance of 0.05), which further reduces the dimensions to 2$D$~\cite{helfrecht2019atomic,gasparotto2021mapping}. The resulting 2$D$ visualization intuitively depicts the protein-ligand structural landscape, where points in close proximity on the map signify structural resemblance.

One can correlate the 2$D$ SOAP UMAP visualization with various properties. For instance, we've integrated Active Sites (AS) and Binding Sites (BS) annotations from the InterPro database~\cite{hunter2009interpro,paysan2023interpro} to provide medicinal chemists and biochemists with an intuitive perspective on the structural intricacies of the CASF dataset. These labels from curated databases are widely used in biology but are often challenging to relate back to atomistic models. Our color-coded low-dimensional map bridges this gap. For a detailed overview of the AS and BS annotations and our extraction methodology from InterPro, please refer to the Supplementary Information (SI).

\section{Results and Discussion}

\subsection{Profiling Structure-Property Relationships}

\begin{figure*}[ht!]
    \centering
    \includegraphics[width=0.8\textwidth]{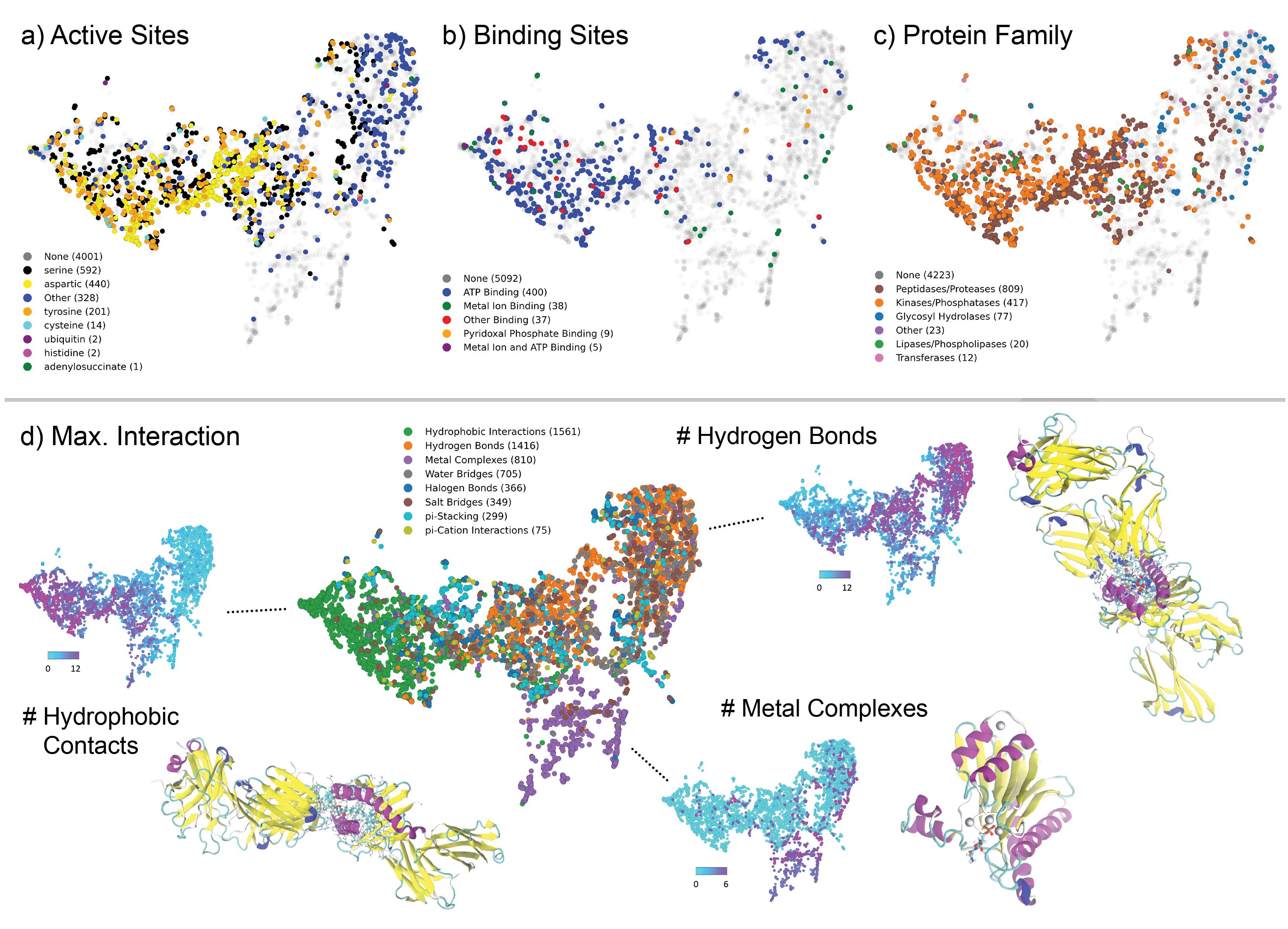}
    \caption{Two-dimensional SOAP UMAP of the RD-2020 set, colored according to different annotations. The top row overlays biological labels extracted from InterPro about active site (a), binding site (b), and (c) protein family classes. The number of annotations extracted for each class is indicated next to the class name. The bottom part shows the projection of PLIP-derived annotations correlated with SOAP map. To provide a compact visualization capturing all interactions, each interaction type is standardized independently, and the dominant interaction type is used to as color in the central map (d). Supplementary maps depict specific interaction types dominant in that area, such as the number of hydrogen bonds, hydrophobic contacts, and metal contacts. Representative snapshots of protein-ligand complex structures are also shown for these specific interactions.}
    \label{fig:casf_total_maps}
\end{figure*}

We generate SOAP vectors to represent the full structural landscape of the RD-2020 dataset using the methods outlined in section \ref{sec:methods-soaps}. We then project the SOAP vectors via UMAP onto a 2$D$ plane and overlay a range of biological and physicochemical annotations (Fig.~\ref{fig:casf_total_maps}). At first glance, the clear structure in the map demonstrates the effectiveness of SOAPs in capturing the high-dimensional features defining protein-ligand complexes. 

\textbf{InterPro Annotations.} Different annotations can be correlated with the 2$D$ SOAP UMAP projection to evaluate and better understand the structure and distribution of the SOAP manifold, offering a unique perspective on potential biases present in the dataset. Here we choose to explore annotations commonly used in medicinal chemistry and biology to stratify proteins, such as active site (AS), binding site (BS), and protein family types.

Amongst the 1,580 complexes annotated with an InterPro AS class, the dominant types are serine (592 occurrences), aspartic (440 occurrences), and tyrosine (201 occurrences). In contrast, only 489 complexes in our dataset are annotated with an InterPro BS class. Notably, the ATP binding site class is overwhelmingly dominant with 400 occurrences, followed by a minor representation of metal ion binding sites. The general skew towards serine and aspartic active sites, or ATP-binding sites, suggests a bias that might limit the diversity of the pharmacological interactions represented in PDBbind. 

The distinction amongst protein family annotations further reveals dominant types like Peptidases/Proteases (809 occurrences) and Kinases/Phosphatases (417 occurrences). The distribution of these protein families notably aligns with the dominant AS and BS types, emphasizing the underlying biochemical roles these proteins are most likely to fulfil. For instance, aspartic active sites are often in the kinase group, whereas ATP BSs are predominantly found in proteases. 

It is interesting to note that ATP binding sites are shifted towards the left side of the map, overlapping with the region characteristic of kinases and peptidases while hydrolases distribute on the opposite side of the map, across a distinct area of the SOAP vector space. 

\textbf{Pharmacophore Annotations.} Human-derived annotations can often be subjective and lean on arbitrary biological functions unrelated to the binding's underlying chemistry. To provide a less biased overview, we overlay the SOAP UMAP projection with properties resulting from the pharmacophore modeling analysis conducted using PLIP. This approach enhances our understanding of how the structural features of different protein-ligand complexes correspond to specific types of protein-ligand interactions, offering a more intuitive viewpoint on the data used in training and testing.

In contrast to biological annotations, we observe a clearer correlation with the 2$D$ SOAP projection (Fig. \ref{fig:casf_total_maps}d). A strong correlation of different pharmacophore-derived properties is evident especially when comparing the spread of each individual property. Structures with high numbers of metal-ion contacts are placed mainly at the bottom of the map. On the other hand, complexes with more hydrophobic contacts are placed towards the left, whereas hydrogen-dominant interactions are placed on the right. In addition, projections derived from compounds with large amounts of hydrogen bonds overlap with those with abundant water bridges. These two observations are chemically sound, considering that hydrophobic contacts are competitive with hydrogen bonds, whereas hydrogen bonds synergise with water bridges. 

The clear clustering of structures in the projection not only confirms the intrinsic validity of the approach, but also provides insights into the functional and chemical nature of the protein that defines different protein-ligand interactions within the dataset. We will exploit these insights to further analyse potential model biases in section \ref{sec:results-generalisability}.

\subsection{CASF-2016 Benchmark}
\label{sec:results-core}
\textbf{Scoring Power.} We start by assessing the scoring power of HydraScreen, \textit{i.e.} the ability to predict the true binding affinity between a ligand and a target protein, for the 285 protein-ligand pairs belonging to the CASF-16 core set. We measure scoring power by comparing the Pearson's correlation coefficient ($r$) and the root mean square error (RMSE) between the true and predicted affinities.

Table \ref{tab:scoring_power} shows the performance of HydraScreen compared to other state-of-the-art methods. We observe that HydraScreen achieves top-tier performance, predicting the affinity of a protein-ligand complex with high accuracy (RMSE=1.15), while also being able to rank compounds according to their affinity ($r$=0.86), a new combined state-of-the-art score on this benchmark. Noteworthy is that most high-performing models in the benchmark such as OnionNet-2 or PointTransformer have been trained with crystal data only, and without the ability to perform pose estimation. However, we believe training with docked poses and using these to distinguish between good and bad poses to be paramount for the robustness and applicablity of HydraScreen. 

At the expense of the added noise, augmenting the training dataset from Refined 2020 to RD-2020 implies a $\sim $150\% increase in training data and boosts the performance considerably. Less noticeable is the performance gain when augmenting from Refined 2016 to 2020 ($\sim$ 40\% increase in training data). This phenomenon is aligned with the commonly reported biases in MLSFs evaluated in CASF-2016: since the similarity between the test and training sets is so high, entries to the benchmark are very likely performing memorisation.

\begin{table*}[ht!]
    \centering
    \caption{Scoring Power of HydraScreen compared to other state-of-the-art methods on the CASF-16 core set. For all MLSFs, training sets are built from the refined (R) and/or the general (G) sets from PDBbind.}
    \begin{tabular}{lrrcc|cc}
        \toprule
        \textbf{Model}                                                            & \textbf{Architecture} & \textbf{Training Set}  & \textbf{Data Type} & \textbf{Pose Estimation}  & \textbf{$r$}  & \textbf{RMSE} \\
        \midrule                                                                                                                   
        Vina \cite{eberhardtAutoDockVinaNew2021, trottAutoDockVinaImproving2010b} & SBDD                  & —                      & —            & —                    & 0.60          & —             \\
        Smina \cite{koesLessonsLearnedEmpirical2013a}                             & SBDD                  & —                      & —            & —                    & 0.55          & —             \\
        MM/GBSA\cite{genhedenMMPBSAMM2015}                                        & —                     & —                      & —             & —                    & 0.65          & —             \\
        Pair \cite{zhuBindingAffinityPrediction2020b}                             & MLP                   & R 2016                 & Crystal            & \XSolid                    & 0.75          & 1.44          \\
        Pafnucy \cite{stepniewska-dziubinskaDevelopmentEvaluationDeep2018a}       & CNN                   & R 2016                 & Crystal            & \XSolid                    & 0.78          & 1.42          \\
        GraphBAR \cite{sonDevelopmentGraphConvolutional2021}                      & GNN                   & R 2016                 & Docked            & \Checkmark                    & 0.78          & 1.41          \\
        GNINA \cite{mcnuttGNINAMolecularDocking2021a}                             & CNN                   & R 2016                 & Docked             & \Checkmark                    & 0.80          & 1.37          \\
        DeepFusion \cite{jonesImprovedProteinLigand2021}                          & CNN + GNN             & R\&G 2016              & Docked            & \XSolid                     & 0.80          & 1.33          \\
        SIGN \cite{liStructureawareInteractiveGraph2021}                          & GNN                   & R 2016                 & Crystal            & \XSolid                    & 0.80          & 1.32          \\
        TopologyNet \cite{cangTopologyNetTopologyBased2017}                       & CNN                   & R 2016                 & Crystal            & \XSolid                    & 0.81          & 1.34          \\
        DeepAtom \cite{liDeepAtomFrameworkProteinLigand2019}                      & CNN                   & R 2016                 & Crystal            & \XSolid                    & 0.81          & 1.32          \\
        FAST \cite{jonesImprovedProteinLigand2021}                                & CNN + GNN             & R 2016                 & Docked            & \Checkmark                     & 0.81          & 1.31          \\
        $K_{deep}$ \cite{jimenezKDEEPProteinLigand2018a}                          & CNN                   & R 2016                 & Crystal            & \XSolid                    & 0.82          & 1.27          \\
        OnionNet \cite{zhengOnionNetMultipleLayerIntermolecularContactBased2019}  & CNN                   & R 2016                 & Crystal            & \XSolid                    & 0.82          & 1.28          \\
		AEScore \cite{meliLearningProteinligandBinding2021a}                      & MLP                   & R 2016                 & Crystal            & \XSolid                     & 0.83          & 1.22          \\
        InteractionGraphNet \cite{jiangInteractionGraphNetNovelEfficient2021}     & GNN                   & R 2016                 & Crystal            & \XSolid                  & 0.84          & 1.22          \\
        PLIG/GATNet \cite{moesserProteinLigandInteractionGraphs2022}              & GNN                   & R 2016                 & Crystal            & \XSolid                   & 0.84          & 1.22          \\
        1D2$D$-CNN \cite{cangRepresentabilityAlgebraicTopology2018a}              & CNN                   & R 2016                 & Crystal            & \XSolid                   & 0.85          & 1.21          \\
        PointTransformer \cite{wangPointCloudbasedDeep2022}                       & CNN + ATT             & R 2016                 & Crystal            & \XSolid                    & 0.85          & 1.19          \\
        OnionNet-2 \cite{wangOnionNet2ConvolutionalNeural2021}                    & CNN                   & R\&G 2019              & Crystal            & \XSolid                  & \textbf{0.86} & 1.16          \\
        \midrule                                                                                                                                                                                                    
                                                                                  & CNN                   & R 2016                 & Docked            & \Checkmark                        & 0.84          & 1.22          \\
        HydraScreen                                                               & CNN                   & R 2020                 & Docked             & \Checkmark                      & 0.84          & 1.21          \\
                                                                                  & CNN                   & RD-2020              & Docked            & \Checkmark                       & \textbf{0.86} & \textbf{1.15} \\
        \bottomrule
    \end{tabular}
    \label{tab:scoring_power}
\end{table*}

\begin{figure}[h]
    \centering
    \includegraphics[width=0.5\textwidth]{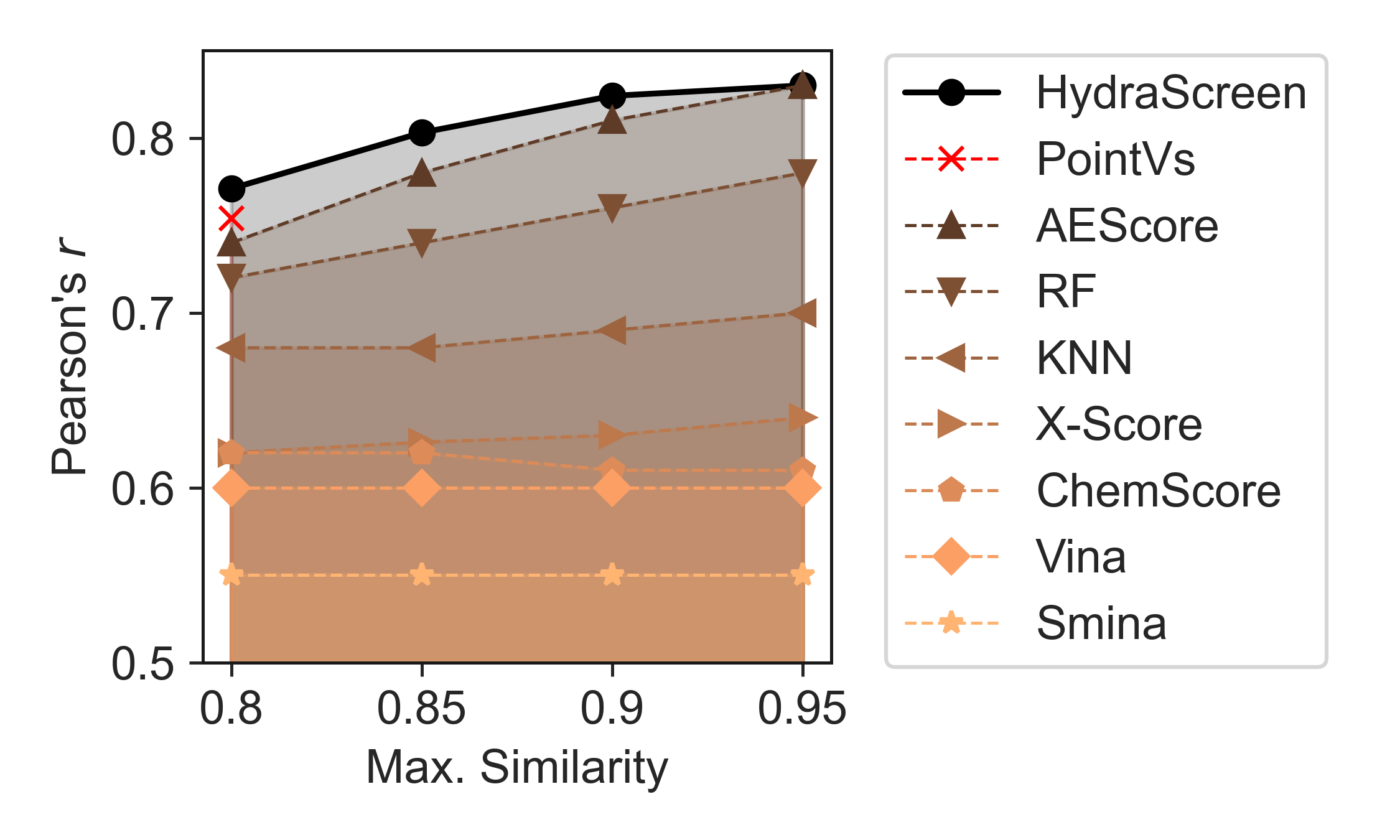}
    \caption{Scoring function performance comparison on the CASF-16 CoreSet benchmark for each of the non-redundant training splits based on Su et al.'s work \cite{suTappingBlackBox2020a}. Note that we include PointVs \cite{scantleburyPointVSMachineLearning2022} as they report values using a very similar test-train split to ours (at a maximum protein sequence and ligand similarity of 0.80).}
    \label{fig:splits}
\end{figure}

\textbf{Generalisation I: Similarity Split.} 
We train HydraScreen with training sets defined by varying the maximum similarity between the training and core sets in line with the procedure outlined in section \ref{sec:datasets}. As shown in Figure \ref{fig:splits}, there exists a clear performance delta relative to the similarity between the training set and the core set. However, in contrast to other MLSFs, HydraScreen shows a consistently lower performance drop within the reduced test-train similarity regimes, suggesting that HydraScreen is able to generalize better to unseen proteins and ligands. 

These results demonstrate a significant improvement over previous methods, which have been shown to be heavily biased towards the training set distribution. In that light, we confirm that the CASF 2016 benchmark is over-confident and is likely not an adequate representation to evaluate if scoring functions have learnt the underlying mechanisms which govern protein-ligand binding, and their corresponding affinity.

\textbf{Screening Power.} We define screening power by the ratio of true binders identified within the top $k\%$ of a compound library. Screening power is a key metric to define the suitability of MLSFs in early-stage drug discovery as it can give a clear estimate of the saving costs translatable to a high throughput screening experiment (by means of prioritizing promising candidates). We evaluate the screening power of HydraScreen by assessing the ability to rank true binders in preference to decoys. To do so, we exploit the CASF-16 screening benchmark, as it consists of 57 different targets with 5 true binders and 280 decoys belonging to each target. We measure screening power by comparing the enrichment factor (EF). EF is computed as the proportion of true binders within the top 1\%, 5\% and 10\% of all 285 compounds ranked according to the scoring function of choice. The overall EF is then calculated by averaging top performance across all 57 targets. 

In order to produce a ranking score for HydraScreen we use the provided docked ensemble of 100 poses per compound-target pair and predict both affinity and PLIE scores. We rank compounds according to the predicted affinity and filter out candidates where PLIE scores are low (PLIE $\leq 0.2$). For fair comparison we ensure that at least 50 candidates are selected per target, reducing the potential over-inflation of the top 1\% metric. 

\begin{figure*}[ht!]
    \begin{subfigure}{0.385\textwidth}  
    \includegraphics[width=\linewidth]{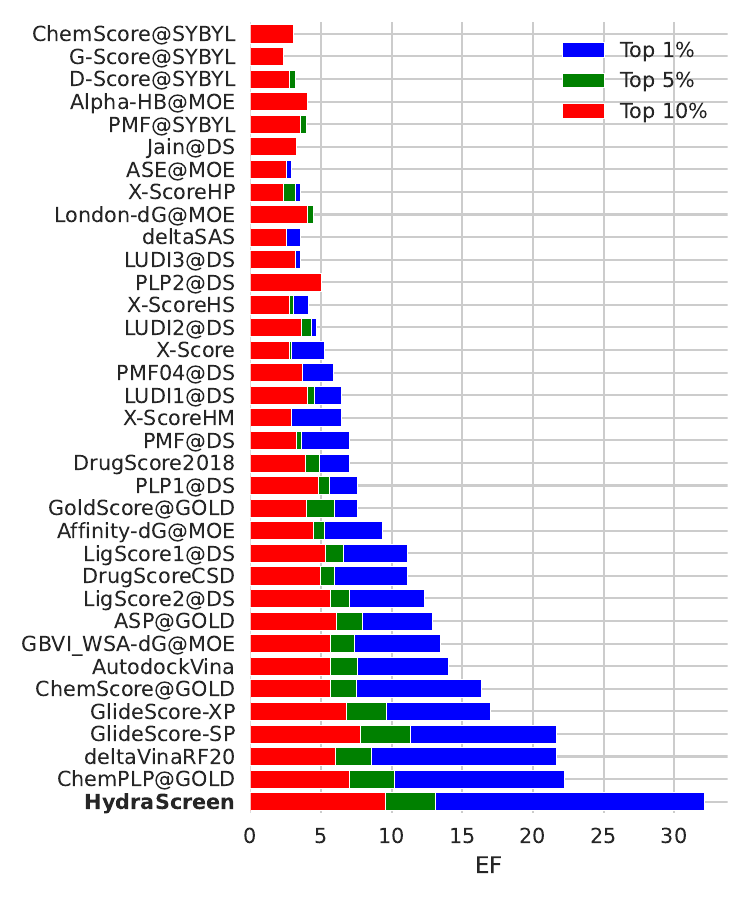}
    \end{subfigure}
    \begin{subfigure}{0.615\textwidth}  
    \includegraphics[width=\linewidth]{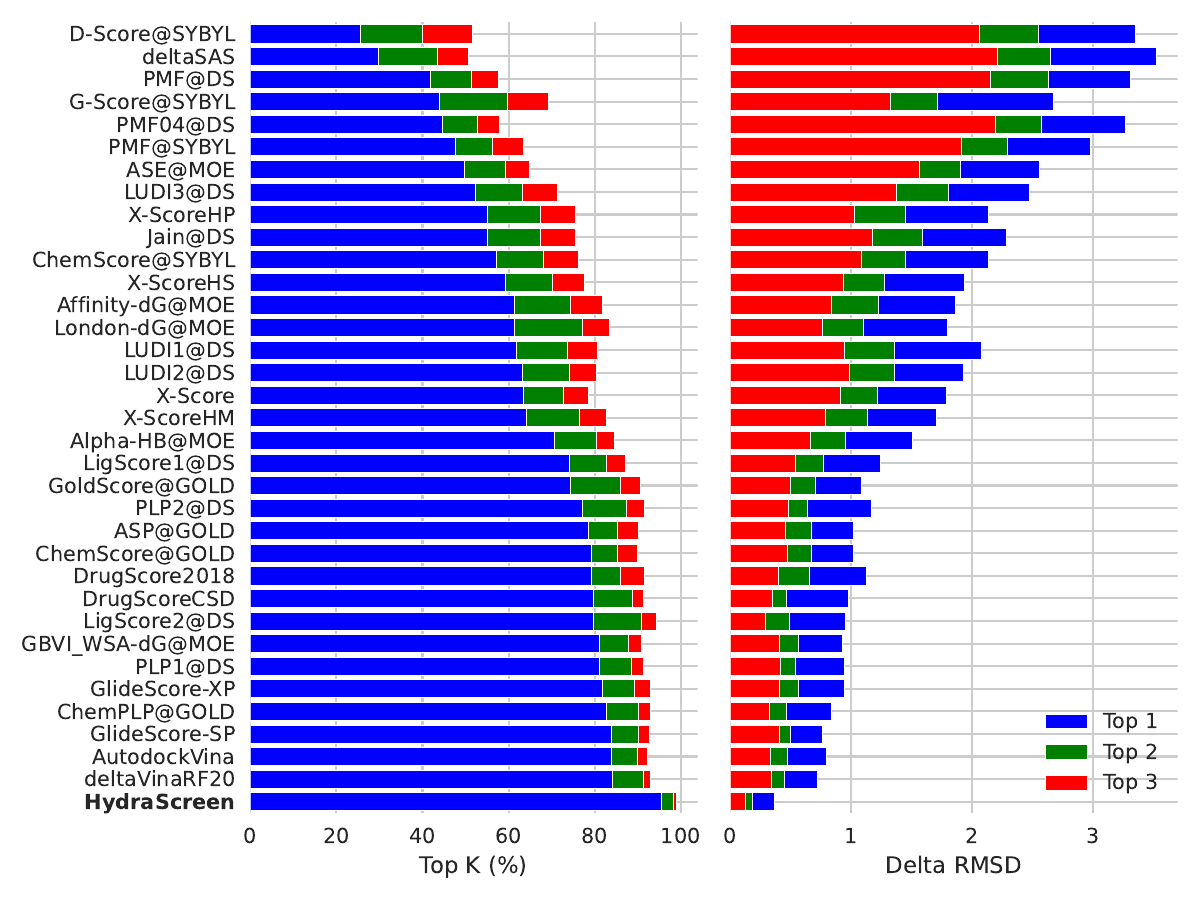}
    \end{subfigure}
    \vspace{-8pt}
    \caption{Comparison between enrichment factors (EF) obtained in the CASF-16 screening benchmark, sorted by highest top 1\% (left). Top $k$ (middle) and $\delta_k^{RMSD}$ (right) evaluated on the CASF docking power benchmark, sorted by highest top 1 score for the top-k metric. Crystal poses were removed to assess performance in a more real-life case scenario.}
    \label{fig:both_figures}
\end{figure*}

We compare our results with the original CASF-16 benchmark in Figure \ref{fig:both_figures} and demonstrate state-of-the-art performance in all three top-k metrics, reaching top 1\% EFs above 32. This result gives strong evidence that HydraScreen can serve as a powerful tool to accelerate virtual screening campaigns: by providing high EFs, one can exploit HydraScreen to select a more potent subset of a virtual library. This not only saves time and experimental costs but also it provides a fail-fast iterative solution to virtual screening.

\textbf{Docking Power.} 
Finding the true binding pose of a ligand is inherently challenging due to the glassy nature of the real Potential Energy Surface governing atomic interactions. This complexity often necessitates severe computational approximations to make the calculations manageable. Additionally, traditional docking methods typically capture only static structural snapshots of the system, focusing solely on enthalpic contributions to free energy. This approach neglects the entropic and dynamic aspects of real experimental systems, which can be critical for understanding system metastability.

In a similar light to screening power, we assess how HydraScreen compares to existing docking methods. The CASF-16 docking power benchmark assesses how often a scoring function can detect a good pose (RMSD $\leq$ 2) from a bad pose (RMSD $>$ 2). For each of the 285 compounds, 100 poses are ranked according to the pose score and the lowest RMSD value within the top $k$ candidates is extracted. We compare the average frequency in which a good pose occurs within the top $k$ and the corresponding average RMSD deviation between the best pose within the top $k$ and the true best pose in the docked ensemble ($\delta$ RMSD). 

We believe $\delta$ RMSD is a more robust representation of true docking power than the traditional top $k$ metric used. As the quality of a docked ensemble will depend on the docking software which generates the docked poses, leading to top $k$ results evaluated across different datasets that are non-translatable or comparable. In addition, top $k$ is based on a subjective threshold and is too punishing: if the lowest RMSD in an ensemble of 100 poses is 1.9~\AA{}, then selecting a pose with an RMSD of 2.1~\AA{} as the best pose should not be equally punishing as selecting a pose with 8~\AA{}. 

\begin{figure*}[ht!]
    \centering
    \includegraphics[width=0.95\textwidth]{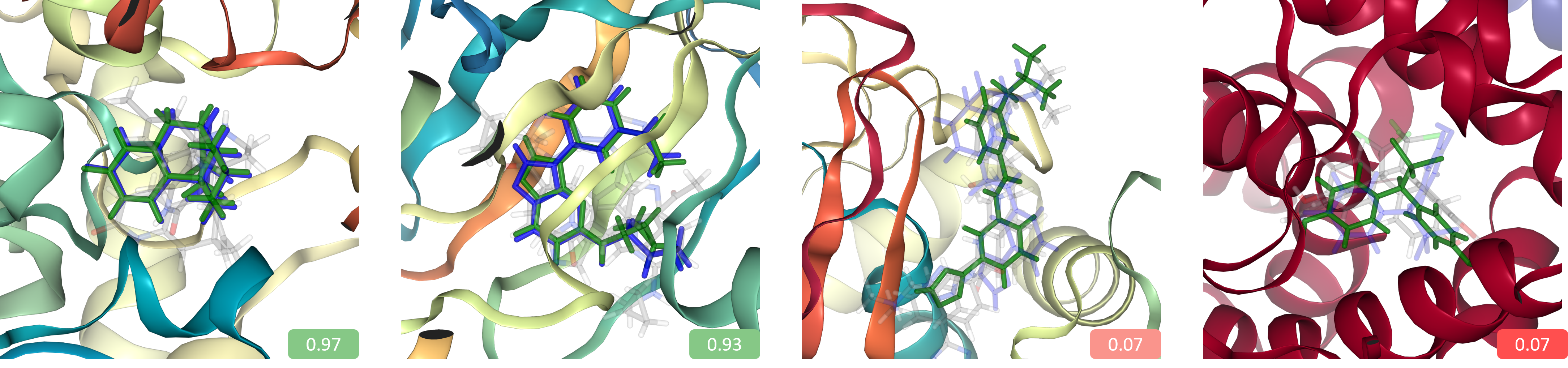}
    \vspace{-4pt}
    \caption{Visualization of a crystal ligand (green), highest scoring pose (blue), and a selection of decoy docked poses (gray) with an opacity proportional to the pose score returned from HydraScreen. The PLIE score is also denoted in the bottom right corner. From left to right, the protein-ligand PDB ids are 1GPN, 2C3I, 4DE2 and 4MGD.}
    \label{fig:docked_pose_scores}
\end{figure*}

A significant increase in docking power is observed with respect to other methods in the docking power benchmark (Fig. \ref{fig:both_figures}). We achieve a near-perfect top 3 accuracy above 99\% and a considerable twofold reduction in $\delta$ RMSD across all top $k$ ranges with respect to the second-best method in the benchmark. In Figure \ref{fig:docked_pose_scores}, we illustrate how, qualitatively, HydraScreen is able to find poses which are very close to the crystal ligand, amongst a range of PDBs from the core set.

We observe that HydraScreen performs well in selecting docked poses that are close to the crystal ligand (1GPN and 2C3I). Moreover, when the docked ensemble has poses that are distant to the crystal structure, the pose score is low as denoted by the low opacity of the highest scoring pose in 4DE2 and 4MGD. This correlates well with the PLIE score which gives an indication as to whether the pose selections are relatively similar to the crystal structure. When the generated docked poses overlap with the native pose, we see a higher PLIE score, and vice-versa. This result is paramount, as it allows the model to discriminate between generated docked poses that are sufficient, and those which are not. An in-depth analysis of the effect on PLIE will be presented in Section~\ref{sec:results-generalisability}.

\subsection{Temporal Split Benchmark}

\label{sec:results-generalisability}
In this section we perform an in-depth analysis of HydraScreen across the rich temporal split defined in section \ref{sec:datasets}. 
To that end, we limit ourselves to training model ensembles over the subset of RD-2020 which overlaps with the PDBbind 2016 refined set, and evaluate them in the held-out-set.

\begin{table*}[ht!]
    \centering
    \begin{tabular}{lrrrrrr}
        \toprule
        \textbf{Model}    & \textbf{Pose Ensemble} & \textbf{$r$} & \textbf{RMSE} & \textbf{Top-1} & \textbf{$\delta_1^{RMSD}$}  & \textbf{AUC} \\
        \midrule
        RF                & Ligand & 0.59                 & 2.17          & - & - & - \\
        LBM               & Ligand & 0.59                 & 2.20          & - & - & - \\
        MPNN              & Ligand & 0.56                 & 2.35          & - & - & - \\
        \midrule
                    & Crystal & 0.49 & - & - & - & - \\
        Smina$^*$       & Docked & 0.51 & - & 0.47 & 1.96 & - \\
                    & Docked \& Crystal & 0.46 & - & 0.46 & 3.03 & - \\
        \midrule
                    & Crystal & 0.49 & - & - & - & - \\
        Vina$^*$        & Docked & 0.50 & - & 0.45 & 1.93 & - \\
                    & Docked \& Crystal & 0.54 & - & 0.41 & 3.32 & - \\
        \midrule
                    & Crystal & 0.59 & 1.57 & - & - & - \\
        GNINA$^{**}$       & Docked & 0.55 & 1.66 & 0.46 & 1.85 & 0.80 \\
                    & Docked \& Crystal & 0.61 & 1.56 & 0.58 & 2.12 & 0.82 \\
        \midrule
                          & Crystal & \textbf{0.71} &\textbf{ 1.30} & - & - & - \\
        HydraScreen       & Docked & \textbf{0.69} & \textbf{1.37} &\textbf{ 0.77} & \textbf{0.71} & \textbf{0.96} \\
                          & Docked \& Crystal & \textbf{0.73} & \textbf{1.29} & \textbf{0.97} & \textbf{0.14} & \textbf{0.97} \\
        \bottomrule
    \end{tabular}
    \caption{
        Comparing scoring and docking power on the PDBbind 2020 - 2016 held-out set. 
        We select protein-ligand pairs with docked ensembles containing at least 20 poses (1085 out of 1529 complexes selected). Computational SBDD (*) and MLSF (**) solutions are used to score the pose ensembles with default parameters.}
    \label{tab:heldout_performance}
\end{table*}

\textbf{Beyond State-of-the-art Performance.}
We extract the key metrics defining scoring and docking power across a series of baselines and summarise them in Table \ref{tab:heldout_performance}. 
We compare our method with traditional force-field energy-based solutions for SBDD (Smina, Vina) as well as MLSFs (GNINA) and ligand-based models (Message Passing Neural Network (MPNN), Random Forest (RF)). 
Details for the latter will be extended in the SI. Models which utilize protein-ligand conformations are evaluated under 3 settings: (i) Only the co-crystallised complex is present; (ii) Only docked poses are present; (iii) Both crystallised and docked complexes are present in the pose ensemble. 

HydraScreen demonstrates superior performance to all baselines in all 3 scenarios, achieving 0.73 $r$ and an RMSE of 1.29 for the affinity prediction of docked ensembles containing the crystal pose.
In comparison to GNINA, our state-of-the-art MLSF baseline, this signifies a $\sim20\%$ increase in correlation between predicted and true affinities and a similar reduction in the RMSE.  
Interestingly, ligand-based approaches which are agnostic to the target achieve a higher correlations than SBDD solutions like Smina and Vina, providing supporting evidence that ligand promiscuity may lead to falsely accurate structure-agnostic models. 
This argument is further evidenced by the difference in RMSE between ligand-based solutions which match correlations with GNINA (although correlation is high, the spread is higher). 
On the other hand, the stark difference in scoring power between ligand-based MLSFs and HydraScreen is significant, with an almost twofold decrease in RMSE.
Indeed, this performance gap may lead us to conclude that key protein-ligand information must be extracted from HydraScreen in order to predict affinity with higher accuracy.

Our approach also achieves state-of-the-art performance in the pose prediction task, reaching near-perfect top-1 of 0.97, insignificant average $\delta_1^{RMSD}$ below 0.14\AA{}, and AUC scores of 0.97 in docked ensembles containing the crystal pose.
Even in the more complex yet practical scenario, where we evaluate on docked poses only, HydraScreen demonstrates substantially higher capabilities, with an average $\delta_1^{RMSD}$ of 0.71\AA{} and a top-1 rate of 0.77. In comparison, these results account for $\sim70\%$ increased success rates and $\sim250\%$ lower $\delta_1^{RMSD}$ RMSDs w.r.t. the second best metric obtained across all models.

As expected, removing crystal poses from the pose ensemble reduces the scoring and docking power of our approach (a similar trend is observed across other baselines). 
On the other hand, by adding docked conformations into the single crystal pose, we observe an increase in performance. 
We believe the reasons for this phenomenon to be twofold. 
First, by adding more poses you invite the model to select more energetically-favourable conformations to that which is provided by the co-crystallised complex.
Second, and most importantly, by predicting the affinity of a protein-ligand complex via a Boltzmann reweighted pose-based energy function, HydraScreen exploits the enriched binding landscape offered by the diverse set of docked poses, deeming it a more robust representation of the model's predicted distribution.

Even though we have evaluated performance in a re-docking scenario where we are guaranteed that the ligand can geometrically fit into its correct pose during simulation, comparison between other methods still provides us with sufficient evidence to claim state-of-the-art performance in the docking power task. To that end, we draw a final observation when comparing the performance metrics obtained when evaluating the core set (section \ref{sec:results-core} and the held-out set (current section). 

Notably, metrics corresponding to scoring power (RMSE and $r$) drop when assessing performance in the held-out set. This is not only true for MLSFs such as HydraScreen or GNINA, but it is also true for ligand-based and parameter-free SBDD models (Vina, Smina), once again demonstrating that the core set is overly optimistic. Indeed, $r$ values above 0.8 are highly unrealistic, especially when considering the inherent noise of the experimental data used to train these models \cite{kalliokoskiComparabilityMixedIC502013}. To that end, this performance drop is expected and therefore does not weaken our claim. 

On the other hand, in contrast to other methods such as Vina which suffer to generalise well to the held-out set, we maintain near-perfect AUC, Top-1 and $\delta_1^{RMSD}$ performance. Maintaining top performance in both evaluation sets indicates that HydraScreen is particularly robust at generalising its discriminative capabilities, as it does not show any particular overfitting nature, deeming it a powerful solution for identifying the true protein-ligand conformation of a ligand binder. 

Finally, we believe the discrepancy between the drop in scoring power and the sustained docking power is also expected, as it is aligned with the difficulty of the task at hand: not only is there more data defining the distribution of high and low RMSD conformations (due to re-docking augmentation), but the task of predicting binding or not is arguably simpler than that of predicting the affinity of the protein-ligand complex. 

\begin{figure*}[ht!]
    \centering
    \begin{subfigure}{0.31\textwidth}  
    \includegraphics[width=\linewidth]{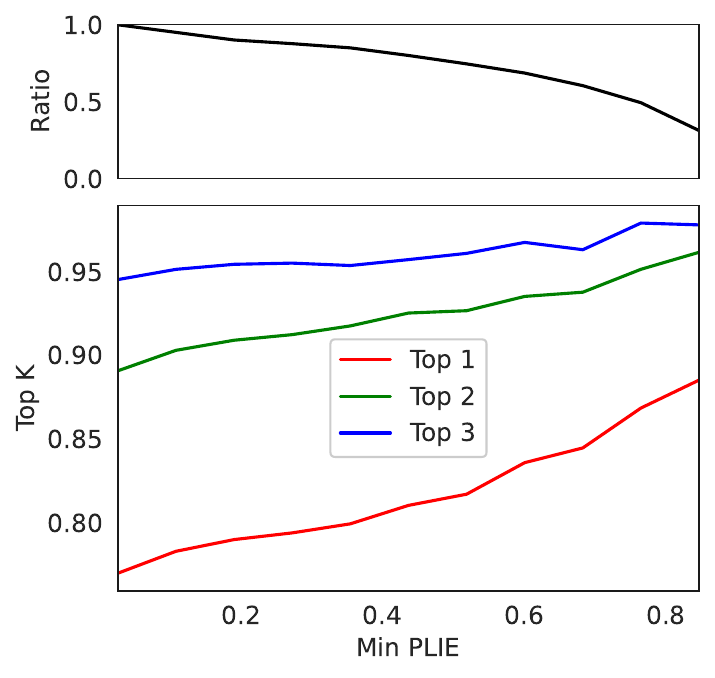}
    \end{subfigure}
    \begin{subfigure}{0.305\textwidth}  
    \includegraphics[width=\linewidth]{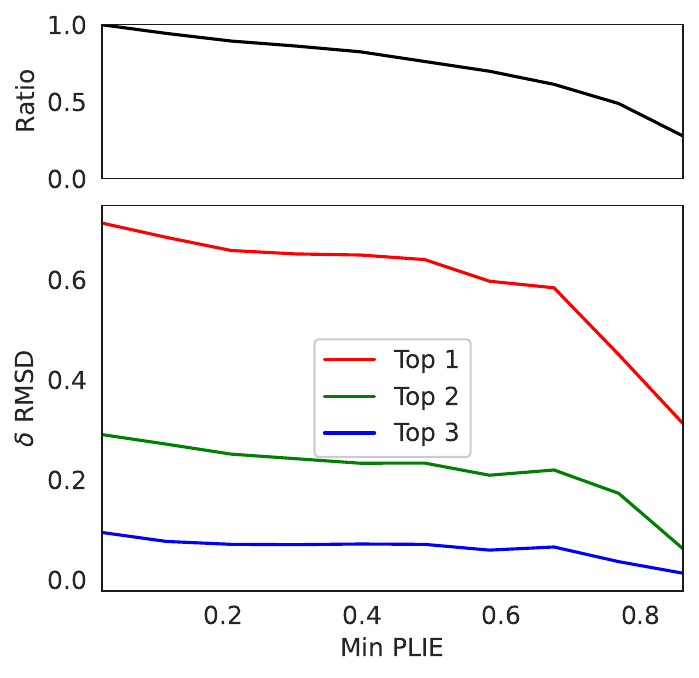}
    \end{subfigure}
    \begin{subfigure}{0.365\textwidth}  
    \includegraphics[width=\linewidth]{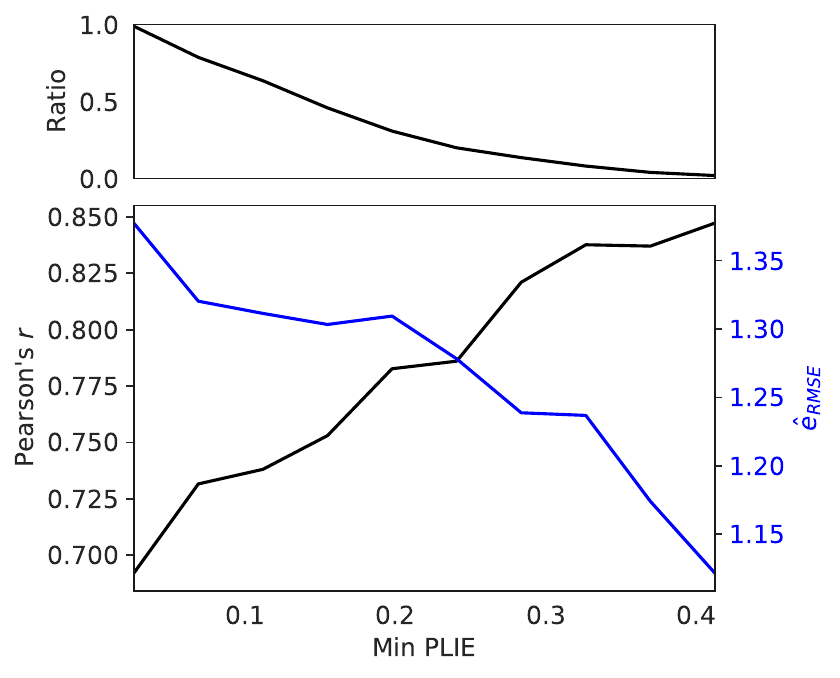}
    \end{subfigure}
    \caption{Assessing the influence of using different minimum thresholds for PLIE score in protein-ligand ensembles extracted from the RD-2020 held-out set. We use docked poses only, excluding the crystal pose from the pose ensemble. On the left and the center, Top-K and $\delta^{RMSD}_k$ performance are measured. Pearson's $r$ and the RMSE between the true and predicted affinities are presented on the left. For every figure, the proportion of complexes included under the minimum PLIE threshold is included for reference. All four metrics improve when pose ensembles with low PLIE scores are discarded.}
    \label{fig:plie-heldout}
\end{figure*}

\textbf{PLIE: A Proxy for Model Confidence.} In line with standard critiques for using poses generated via docking for affinity and pose prediction tasks, we devised PLIE score (Eq. \ref{eq:plie}) to better understand if HydraScreen can estimate when a docked ensemble is inadequate or insufficient. As we demonstrated in Figure \ref{fig:docked_pose_scores}, pose ensembles with higher PLIE scores often lead to docked poses closer to the true pose. Hence, to clarify its effect, we split the evaluation set according to a minimum PLIE score threshold and look at the effect of increasing such threshold across the protein-ligand pose ensembles derived from the held-out set. The corresponding results are reported in Figure \ref{fig:plie-heldout}.

For both binding pose and affinity scoring tasks, increasing the PLIE threshold monotonically reduces the prediction error. 
The inverse temperature, $\beta$, is utilized in Boltzmann reweighting to influence the entropic contribution tied to the structural ensemble. We use $\beta$=1/2 to assess affinity scores based on the PLIE score (in a higher $T$, or more entropic scenario), and $\beta$=8 (in a lower $T$, or more enthalpic regime) to rank pose ensembles for pose prediction. Details on the applicability and sensitivity of $\beta$ are discussed more in detail in the SI. In Figure \ref{fig:plie-heldout} we observe how $r$ values increase from 0.69 to 0.87 and RMSE values decrease from 1.37 to 1.01. Similarly, Top-1 accuracies increase from 0.77 to 0.92, and $\delta_1^{RMSD}$ values reduce from 0.71 to 0.27. Noteworthy is that although PLIE is directly indicative of the predicted pose distribution (Eq. \ref{eq:plie}), it is not formally bound to the affinity distribution. However, the clear increase in prediction accuracy demonstrates the otherwise desired entanglement between the quality of the pose ensemble and the accuracy of HydraScreen's predicted affinity. 

The monotonically increasing accuracy validates that PLIE provides another layer of robustness for HydraScreen, allowing it to flag the notoriously poor poses often generated with traditional docking pipelines which could be discarded to produce more accurate and robust predictions. To that end, we believe PLIE could be used in drug discovery campaigns to: (i) discard ligands in HTS stages which fail to adequately dock into a pocket (low $\beta$), and/or (ii) support existing pose generation tools as an oracle, with the aim of providing additional poses until a certain PLIE criteria is met (high $\beta$).

\textbf{Interaction Profiling.} To assess the interpretability and the potential biases of our methodology, we examined the SOAP maps generated in section \ref{sec:methods-soaps} in the lens of HydraScreen's performance over the held-out set. 

\begin{figure}[ht!]
\centering
\includegraphics[width=0.45\textwidth]{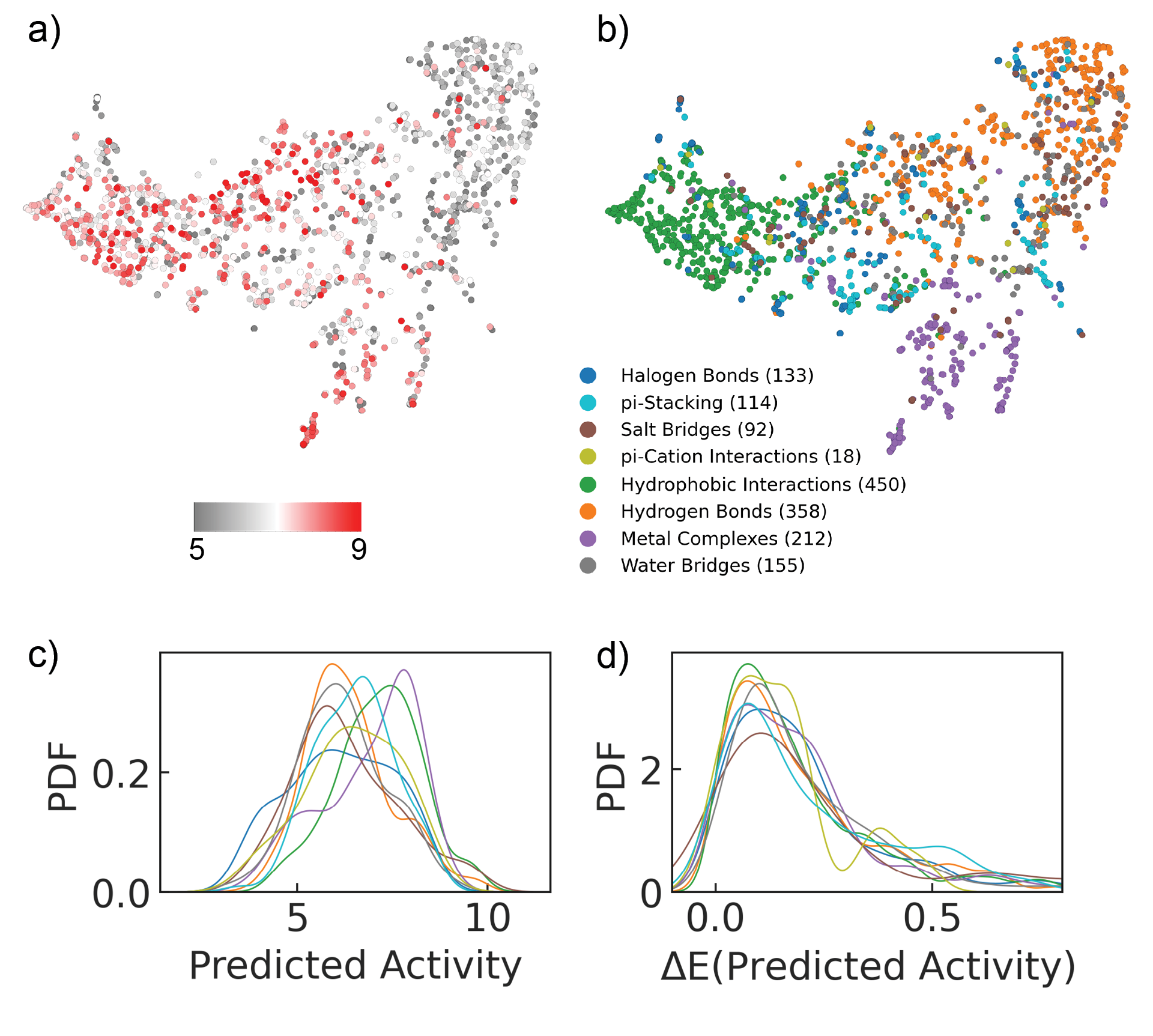}
\caption{Two-dimensional SOAP UMAP depicting HydraScreen's performance on the held-out set in predicting affinities for crystal poses (a). High-affinity complexes are predominantly located on the left and bottom side of the map, area dominated marked by hydrophobic interactions as shown in panel b. Panel c shows the distribution of predicted affinities stratified by dominant interaction types, as introduced in the clusters in panel b. Similarly, panel d shows the error in the affinity prediction.}
\label{fig:heldout_maps_hydraperf}
\end{figure}

In Figure \ref{fig:heldout_maps_hydraperf}a we overlay a heatmap proportional to the predicted affinity. High-affinity complexes predominantly cluster on the left (hydrophobic rich) and bottom (metal-ions rich) sides of the map. Differences in affinity distributions, stratified by interaction-types, are further evidenced in Figure~\ref{fig:heldout_maps_hydraperf}c, where a clear shift in the predicted affinity modes appears across types of predominant interaction classes. Notably, structures dominated by hydrogen bonds tend to fall at lower affinities compared to those dominated by hydrophobic or metal-ion interactions. Interestingly, the distribution of halogen bonds is more dispersed than that of HBs, reflecting the looser and less directional nature of halogen bonds in determining ligand affinity. 

When instead we analyse the normalized error in affinity prediction (Fig.~\ref{fig:heldout_maps_hydraperf}d), the lack of distributional shift suggests that HydraScreen does not exhibit any particular bias toward interaction-dominant interactions. In contrast to our initial intuition, we expected higher error margins in complexes with large amounts of metal-ion and water bridge contacts, as these were not included as part of the model's feature representation layer. However, given the lack of specific bias towards these interaction types, we believe that HydraScreen has implicitly learnt to "fill in the gap" amongst the otherwise relevant species which are removed from the pocket during training.

It is also worth noting that structures containing hard-to-model or structurally loose chemical contacts such as halogen bonds, pi-cation, water bridges, and metal-ion have distributions with a broader right tail. We argue this could potentially be due to the fact that both experimental data and theoretical tools used for fitting crystal structures present intrinsic biases that may lead to noisier structures that are harder to learn for the model. 

Collectively, the lack of clear strong biases when testing Hydrascreen on out-of-sample data is crucial, as it offers additional proof that the model possesses significant generalizing capability across diverse protein-ligand complexes.

\section{Conclusion}

In this study, we introduced HydraScreen, a state-of-the-art 3$D$ CNN-based approach that seamlessly integrates the prediction of both ligand pose and binding affinity with state-of-the-art accuracy.
This comprehensive framework distinguishes itself from existing methodologies which often focus solely on either pose or affinity prediction, highlighting HydraScreen's versatility in the landscape of structure-based drug discovery.

Through meticulous analysis, we demonstrate HydraScreen's commendable generalization power. Notably, within the confines of the established CASF 16 benchmark, its efficacy becomes particularly conspicuous when confronted with a marked diminution in training-test similarity or when assessed against an enriched test set defined via temporal partitioning. In these contexts, we manifest the potential of crafting robust representations of protein-ligand interplays utilising PLIP-enhanced SOAP vectors. Our findings corroborate that HydraScreen exhibits no discernible inclination towards any specific interaction-dominant binding. Furthermore, we affirm the sagacity of our advanced technique, which harnesses Boltzmann reweighted pose ensembles to refine predicted affinities. Concurrently, we underscore the salience of PLIE, an innovative metric introduced to appraise the fidelity of docked ensembles and gauge the confidence vested in the model.

We divide our future directions into two main avenues. The first encompasses the training paradigm and architecture. In addition to creating invariant architectures with adequate inductive biases to efficiently abstract atomic interactions and environments (i.e. by properly incorporating symmetry imposed by physical laws), we believe that a key next step in designing robust MLSFs would involve shifting the training paradigm from supervised to semi-supervised learning. An example of such a paradigm would involve contrasting learning across sets of protein-ligand complexes.

A second, equally important direction involves creating larger, more inclusive datasets by exploiting self-supervision to bootstrap extensive unlabelled protein-ligand annotations. Moreover, the inclusion of ``negative data'' and adequate metrics to define true proximity to the native pose could be explored. We believe that current attempts to include decoys in training datasets do not necessarily improve the generalisability of MLSFs, as the metrics defining proximity are not well established: RMSD between different molecules is not chemically sound.

Overall, our state-of-the-art performance against traditional structure-based and machine-learning methods not only validates the robustness of our method but also reinforces the suitability of AI-based solutions in this field. We encourage users to use the free HydraScreen web interface available at \href{https://hydrascreen.ro5.ai} across different stages of drug discovery, including virtual screening and lead optimization.

\section*{Data Availability}
We release the docked poses as SDF files from our RD-2020 dataset available at: \href{https://www.Ro5.ai/}{https://www.Ro5.ai}.

\section*{Acknowledgements}
We are grateful to Dr. Sarah Flatters, Orestis Bastas, and Charles Dazler Knuff for enriching our scientific discussions. Special thanks to Siim Schults and Dainius Salskaukas not only for their pivotal role in the development of our web interface but also for their invaluable support in streamlining code standards and facilitating the code's deployment to the cloud. 

\section*{Author Contributions}
A.P. led the implementation and design of Hydrascreen. A.P., P.G. and H.A.A. wrote the first version of the manuscript. P.G. and A.P. implemented the interaction profiling and SOAP analysis. A.P. and P.G. discussed and designed the Hydrascreen's architecture and trained the models. A.P. and H.A.A. performed the docking experiments. A.P., P.G. and H.A.A. designed the experiments and processed the results, with R.T. providing valuable contributions. All authors discussed the theory and results, and revised the manuscript.

\bibliography{main}%
\bibliographystyle{abbrv}
\end{document}